\documentclass[pra,twocolumn,showpacs,preprintnumbers,amsmath,amssymb,superscriptaddress]{revtex4}
\pdfoutput=1
\usepackage{color}
\usepackage{graphicx}% Include figure files
\usepackage{dcolumn}% Align table columns on decimal point
\usepackage{bm}% bold math

\usepackage[T1]{fontenc}
\newcommand{\ket}[1]{| #1 \rangle}

\newcommand{\proj}[1]{| #1 \rangle \langle #1 |}
\newcommand{\al}{\alpha}

\newcommand{\op}[1]{ \hat{\sigma}_{#1} }

\begin{document}

\title{Nonlocality activation in entanglement swapping chains}

\author{Waldemar K\l{}obus}%
%\email{ald@amu.edu.pl}
\affiliation{Faculty of Physics,
 Adam Mickiewicz University,
Umultowska 85, 61-614 Pozna\'{n}, Poland.
}

\author{Wies\l{}aw Laskowski}%
%\email{wieslaw.laskowski@univ.gda.pl}
\affiliation{Institute of Theoretical Physics and Astrophysics,
University of Gda\'{n}sk, 80-952 Gda\'{n}sk, Poland.
}

\author{Marcin Markiewicz}%
%\email{wieslaw.laskowski@univ.gda.pl}
\affiliation{Institute of Theoretical Physics and Astrophysics,
University of Gda\'{n}sk, 80-952 Gda\'{n}sk, Poland.
}

\author{Andrzej Grudka}%
%\email{agie@amu.edu.pl}
\affiliation{Faculty of Physics,
 Adam Mickiewicz University,
Umultowska 85, 61-614 Pozna\'{n}, Poland.
}

\date{\today}

\begin{abstract}
We consider multiple entanglement swappings performed on a chain of bipartite states. Each state does not violate CHSH inequality. We show that before some critical number of entanglement swappings is achieved the output state does not violate this inequality either. However, if this number is achieved then for some results of Bell measurements obtained in the protocol of entanglement swapping the output state violates CHSH inequality. Moreover, we show that for different states we have different critical numbers for which CHSH inequality is activated.

\end{abstract}

\pacs{03.65.Ud  05.50.+q}% PACS, the Physics and Astronomy

\maketitle

%\section{Introduction}\label{sec1}

Nonlocal correlations between outcomes of measurements performed on separated subsystems, manifested in the violation of Bell inequalities, is the most profound feature that characterizes quantum description in opposition to the classical one. Such typically quantum correlations may be found of practical interest in the field of information processing and communication protocols which in many instances outperforms their classical counterparts.
Nonlocal correlations between outcomes of local measurements performed on separated particles can be obtained only if particles are entangled. However, this is not sufficient condition, since there exist entangled states, which admit hidden variable model and no Bell inequality can be violated with them \cite{RW1}. Popescu \cite{SP1} and Gisin \cite{NG1} showed that if one or two parties perform measurements on such states then the post measurement state can violate CHSH inequality.

Even more interesting situation arises in the process of entanglement swapping \cite{MZ1, MZ2}. Let us now suppose that Alice and Bob as well as Bob and Charlie share two-qubit entangled states which do not violate CHSH inequality. It was shown in \cite{WMGC} that if Bob performs Bell measurement on his qubits - one from a state which he shares with Alice and one from a state which he shares with Charlie then for some initial states and for two results of his measurement the final state shared by Alice and Charlie violates CHSH inequality.
In a recent paper \cite{CRS} a similar scenario was presented in the context of nonlocality tests.
Although we call this effect activation of nonlocality, it has to be distinguished from activation of nonlocality as presented in Ref. \cite{Ver}, where it is achieved between two parties sharing two copies of a bipartite state which does not violate CHSH inequality.

In the present paper we consider multiple entanglement swappings performed on a chain of bipartite states and show that before some critical number of entanglement swappings is performed the output state does not violate CHSH inequality. However, if we perform sufficiently large number of entanglement swappings equal to some critical value then nonlocality is activated. Moreover this process of activation can be performed
in principle even in the case when states initially possessed by the parties are very weakly entangled.

%\section{SEC II}\label{sec2}

\begin{figure}
  \includegraphics[width=0.45\textwidth]{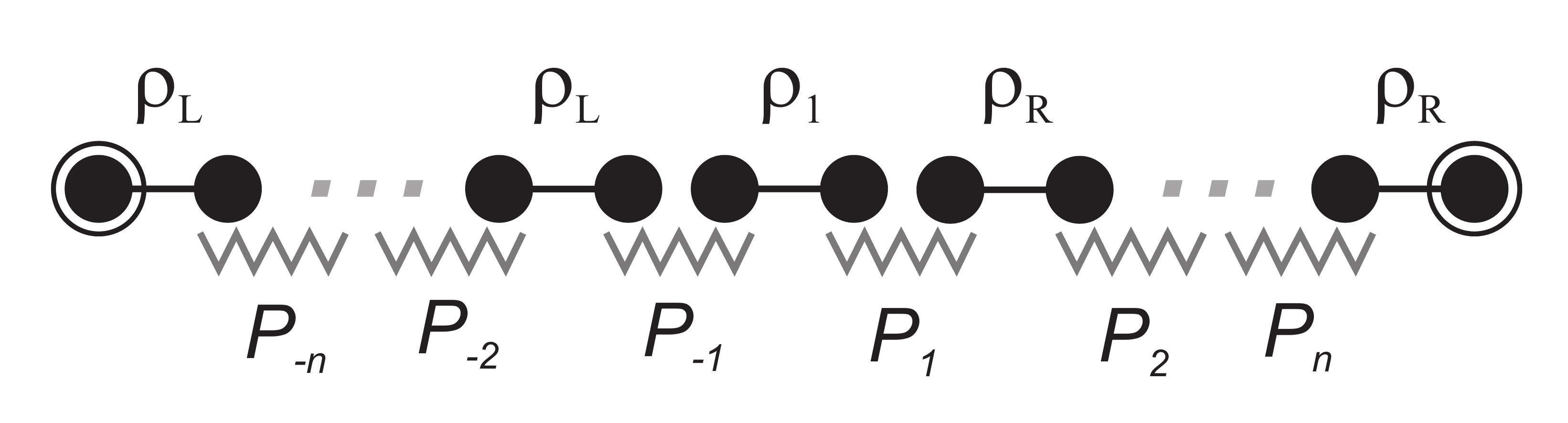}\\
  \caption{Chain of entanglement swappings. Each pair of parties $P_{-i-1}$, $P_{-i}$ shares a state $\rho_L$ and each pair of parties $P_i$, $P_{i+1}$ $(1\leq i \leq N-1)$ shares a state $\rho_R$. The state $\rho_1$ is shared by parties $P_{-1}$ and $P_{1}$. Each party performs Bell measurement on his qubits. }\label{chain}
\end{figure}

Let us consider a total amount of $2N-1$ two-qubit states distributed among $2N$ parties $P_{-N}$, $P_{-N+1}$, ..., $P_{N}$ in chain (see Fig. \ref{chain}) in such a way that each party of a pair $P_i$, $P_{i+1}$ $(-N\leq i \leq -2)$ share a state $\rho_L$ and similarly each party of a pair $P_i$, $P_{i+1}$ $(1\leq i \leq N-1)$ share a state $\rho_R$, and additionally the parties $P_{-1}$, $P_1$ share a state $\rho_1$, where
\begin{eqnarray}
\rho_L = p\proj{\Psi_L} + (1-p)\proj{00},
\end{eqnarray}
with
\begin{eqnarray}
\ket{\Psi_L} = \cos \al \ket{01} + \sin \al \ket{10},
\end{eqnarray}
and similarly
\begin{eqnarray}\label{roR}
\rho_R = p\proj{\Psi_R} + (1-p)\proj{00},
\end{eqnarray}
with
\begin{eqnarray}
\ket{\Psi_R} = \sin \al \ket{01} + \cos \al \ket{10},
\end{eqnarray}
whereas
\begin{eqnarray}
\rho_1 = p_1\proj{\Psi^+} + (1-p_1)\proj{00},
\end{eqnarray}
with
\begin{eqnarray}
\ket{\Psi^\pm} = \frac{1}{\sqrt2} ( \ket{01} \pm \ket{10}).
\end{eqnarray}

\begin{figure}
  \includegraphics[width=0.4\textwidth]{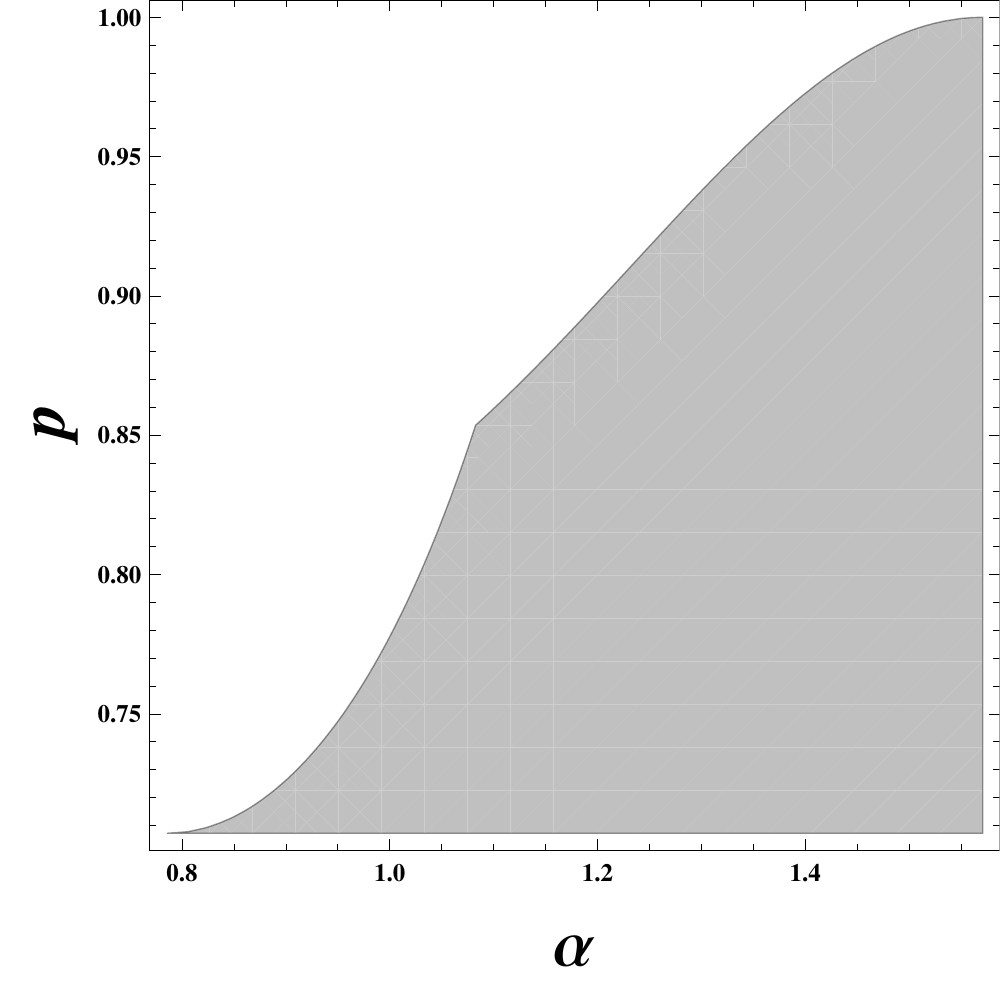}\\
  \caption{The states $\rho_L$, $\rho_R$ which do not violate CSHS inequality (shaded region).}\label{roRL}
\end{figure}

At this stage we are interested solely in the class of states that do not exhibit a violation of CHSH inequality. Let us denote by $\lambda_i$ the eigenvalues of $R^T R$, where $R_{ij} = \textrm{Tr}[(\op{i} \otimes \op{j}) \rho] $ ($\op{i}$ - standard Pauli matrices) then the state $\rho$ does not violate CHSH inequality if for each pair of eigenvalues the following condition holds $\sqrt{\lambda_i + \lambda_j} < 1$ \cite{Hbell}. Using the above criterion it can be verified that if the parameters satisfy the following conditions
\begin{eqnarray}
& p_1 \leq \frac{1}{\sqrt2} \approx 0.707, \\
& \max_{p,\al} \left[ 2 p^2 \sin2\al ,\,\, 1-4p + \frac{p^2}{2}(9-\cos4\al) \right] \leq 1,
\end{eqnarray}
(Fig. \ref{roRL} displays the corresponding range of given parameters in $\al - p$ plane) then each of the states $\rho_L$, $\rho_R$ and $\rho_1$ does not violate CHSH inequality.

Let us now describe the procedure of entanglement swapping that we will use below. First, the parties $P_{-1}$ and $P_1$ perform Bell measurement on their qubits and hence, from the chain of states $\rho_L \otimes \rho_1 \otimes \rho_R$ they produce some output state $\rho_2$ which is shared by the parties $P_{-2}$ and $P_2$. Next, the parties $P_{-2}$ and $P_2$ perform Bell measurement on their qubits and from the chain of states $\rho_L \otimes \rho_2 \otimes \rho_R$ they produce some output state $\rho_3$ which is shared by the parties $P_{-3}$ and $P_3$ and so on. We note that the analysis is independent of the length of the chain since the initial state is repeatedly affected in the same way. The exact form of the output state depends solely on results of all Bell measurements. If all parties $P_{-k+1}$,..., $P_{k-1}$ obtain $\ket{\Psi^\pm}$ as results of their measurements and the party $P_{k}$ corrects the phase then the parties $P_{-k}$,..., $P_{k}$ will share a state

\begin{eqnarray}\label{ropon}
\rho_k = p_k\proj{\Psi^+} + (1-p_k)\proj{00},
\end{eqnarray}
where
\begin{eqnarray}
p_k  = \left[ \frac{\cot^{2(k-1)}\al}{p_1} + (1-\cot^{2(k-1)}\al) \left( \frac{p-1}{p \cos 2\al} + 1\right) \right] ^{-1}.
\end{eqnarray}
One can see that entanglement swapping changes the ratio of the maximally entangled state $\ket{\Psi^\pm}$ to noise.

Because for states of the form of (\ref{ropon}) the necessary and sufficient condition for violation of CHSH inequality is
\begin{eqnarray}
p_k > \frac{1}{\sqrt2}
\end{eqnarray}
and $p_k$ increases with $k$ for some $p$ and $\al$,
we can transform a chain of some initial states $\rho_L$, $\rho_1$ $\rho_R$ (i.e. with some particular values of $p$ and $\al$) which do not violate CHSH inequality into a state $\rho_k$ which violates this inequality by performing sufficiently large number of entanglement swappings.
In this sense nonlocality can be activated. However the probability of performing $m$ entanglement swappings with measurements outcomes $\ket{\Psi^\pm}$ decreases exponentially with $m$.

\begin{figure}
  \includegraphics[width=0.4\textwidth]{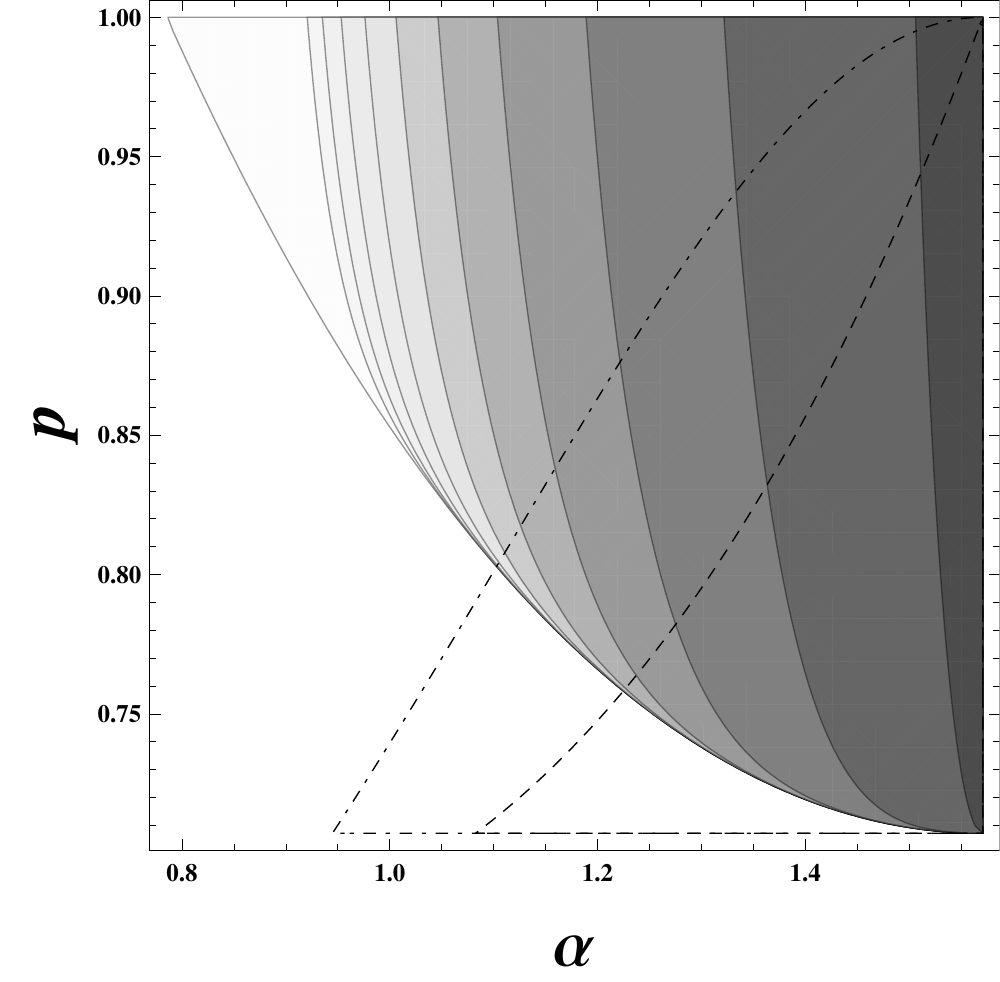}\\
  \caption{The states $\rho_L$ and $\rho_R$ for which CHSH inequality is activated after $n$ entanglement swappings when all measurements outcomes are $\ket{\Psi^\pm}$ for $n=2,4,...,20$ and $n\rightarrow\infty$ (shaded regions from right to left) and $p_1=0.01$.
 The states for which the measurement outcome $\ket{\Phi^\pm}$ gives rise to separable state when the measurement is performed on a state $\rho_R \otimes \rho_R$ ($\rho_1 \otimes \rho_R$) are below dashed (dash-dotted) line for $p_1 = \frac{1}{\sqrt2}$.}\label{ropo}
\end{figure}

In Fig. \ref{ropo} we present the range of parameters of states $\rho_L$ and $\rho_R$ for which nonlocality is activated after several entanglement swappings (with the use of the initial state $\rho_1$ with an arbitrary $p_1 = 0.01$). One can see that parts of the region corresponding to states which do not violate CHSH inequality (see Fig. \ref{roRL}) contain states for which nonlocality is activated after performing sufficiently large number of entanglement swappings, i.e., it happens that $m$ entanglement swappings are insufficient to obtain CHSH violation, a property that is available only after additional $2$ entanglement swappings.

%\begin{figure}
%  \includegraphics[width=0.4\textwidth]{comb2.pdf}\\
%  \caption{The states for which the measurement outcome $\ket{\Phi^\pm}$ gives rise to separable state when the measurement is performed on a state %$\rho_R \otimes \rho_R$ (the region below dashed line) or on a state $\rho_1 \otimes \rho_R$ for $p_1 = \frac{1}{\sqrt2}$ (the region below %dash-dotted line) in comparison with states which activate nonlocality when all measurements outcomes are $\ket{\Psi^\pm}$.}\label{peho}
%\end{figure}

\begin{figure}
  \includegraphics[width=0.4\textwidth]{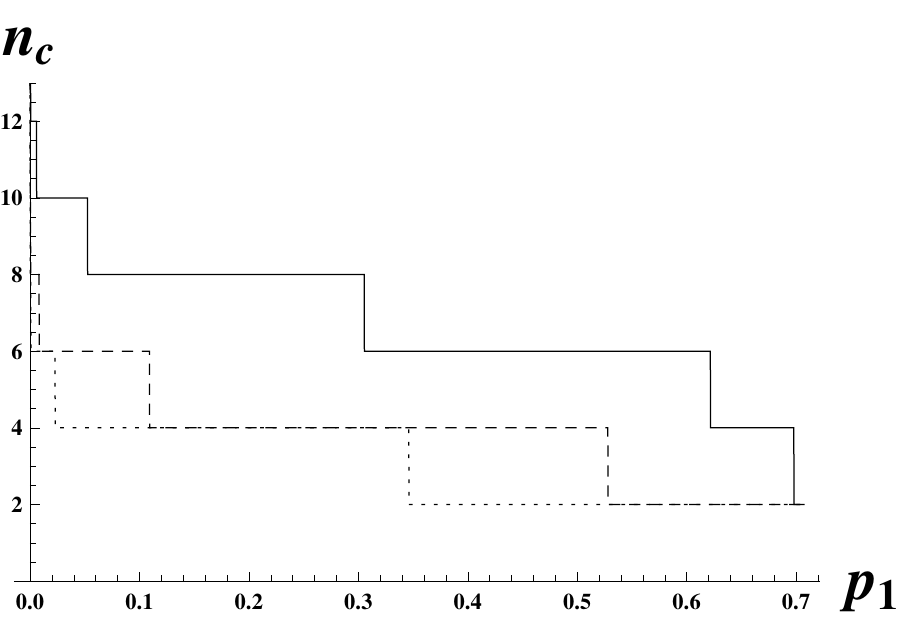}\\
  \caption{Critical number of measurements $n_c$ with all measurements outcomes $\ket{\Psi^\pm}$  needed for activation of nonlocality for several different initial states $\rho_L$ and $\rho_R$ ($p=0.75$) with $\alpha=\frac{20}{25}\pi$ (solid line), $\alpha=\frac{21}{25}\pi$ (dashed line) and $\alpha=\frac{22}{25}\pi$ (dotted line).}\label{nc}
\end{figure}

In order to show explicitly that the number of measurements with all measurements outcomes $\ket{\Psi^\pm}$ is of primary importance we may evaluate the critical number $n_c$ of entanglement swappings needed to attain the CHSH violation while using different states $\rho_1$, $\rho_L$ and $\rho_R$. We see that for some states $\rho_L$ and $\rho_R$ it is possible to achieve the CHSH violation for any states of the form $\rho_1$ even for arbitrarily large amount of initial noise. Fig. \ref{nc}  illustrates the critical number $n_c$ for some initial states $\rho_1$, $\rho_L$ and $\rho_R$ in the chain.

%\begin{table}\label{tab}
%\begin{equation*}
%\begin{array}{lccc}
%\hline \hline
%p_1  & p    & \al         & n_c   \\ \hline
%0.01 & 0.75 & \frac25 \pi & 10   \\
%0.5  & 0.75 & \frac25 \pi & 6   \\
%0.7  & 0.75 & \frac25 \pi & 2   \\
%0.01 & 0.75 & \frac49 \pi & 6   \\
%0.5  & 0.75 & \frac49 \pi & 2   \\
%0.7  & 0.75 & \frac49 \pi & 2   \\ \hline \hline
%\end{array}
%\end{equation*}
%\caption{Critical number of measurements with all measurements outcomes $\ket{\Psi^\pm}$  needed for activation of nonlocality for several different initial states $\rho_L$ and $\rho_R$ and $p_1=0.01$.}
%\end{table}

%\section{SEC III}\label{sec3}

Let us now suppose that the parties perform smaller number of entanglement swappings with other measurement outcomes and check if the outcome state violates CHSH inequality. We performed numerical calculations for the cases where (i) party $P_1$ performs entanglement swapping, (ii) party $P_{-1}$ performs entanglement swapping, (iii) parties $P_{-1}$ and  $P_{1}$ perform entanglement swappings and so on up to the case where parties $P_{-3}, P_{-2}, P_{-1}, P_{1}, P_{2}$ and  $P_{3}$ perform entanglement swappings and all possible configurations of measurements outcomes. We found that if we cannot activate nonlocality between parties $P_{-k}$ and $P_{l}$ if the parties $P_{-k+1},...,P_{l-1}$ obtain $\ket{\Psi^\pm}$ as results of Bell measurements then we also cannot activate nonlocality for any other configuration of measurements outcomes. Unfortunately the number of configurations of measurements outcomes grows exponentially with the number of parties which perform entanglement swapping which makes numerical calculations inefficient. However, we can prove for some initial states $\rho_L$, $\rho_1$ and $\rho_R$ (for which we can activate nonlocality if $n_c$ parties obtain $\ket{\Psi^\pm}$ as results of their measurements) and arbitrary $n_c$, we cannot activate nonlocality between any parties by performing smaller number of entanglement swappings. We do this by showing that

(\textit{i}) we cannot activate nonlocality between parties $P_{-k}$ and $P_{k}$ if at least one of the parties $P_{-k+1},...,P_{k-1}$ obtains $\ket{\Phi^\pm}$ instead of $\ket{\Psi^\pm}$ as a result of his measurement, where
\begin{eqnarray}
\ket{\Phi^\pm} = \frac{1}{\sqrt2} ( \ket{00} \pm \ket{11});
\end{eqnarray}

(\textit{ii}) we cannot activate nonlocality between party $P_{k}$ ($P_{-k-n}$) and $P_{k+n}$ ($P_{-k}$), where $k \geq 1$ and $n \geq 2$;

(\textit{iii}) we cannot activate nonlocality between party $P_{-k}$ ($P_{k}$)  and $P_{k+n}$ ($P_{-k-n}$), where $k, n \geq 1$.

(\textit{i}) It is clear that if we substitute one of the entangled states $\rho_L$ or $\rho_R$ in the chain of states by some separable state $\rho_{S}$ then any procedure of entanglement swapping involving this state cannot give rise to activation of nonlocality. It suffices to consider only one part of the chain consisting of states $\rho_R$ shared by parties $P_1$, ..., $P_k$ for $k>1$ (by the symmetry of states $\rho_L$ and $\rho_R$ the argument is the same for the second part of the chain consisting of states $\rho_L$ shared by parties $P_{-1}$, ..., $P_{-k}$). Let us suppose, that the party $P_i$ ($1 < i < k$) obtains $\ket{\Phi^\pm}$ as an outcome of Bell measurement. In such a case parties $P_{i-1}$ and $P_{i+1}$ will share some two-qubit state $\rho^{\Phi^\pm}_{RR}$. Using Peres-Horodecki separability criterion for density matrices \cite{AP, Hsep} it can be shown that for some $p$ and $\al$  $\rho^\Phi_{RR}$ is separable.
Similarly if the party $P_1$, who shares a state $\rho_1$ with the party $P_{-1}$ and a state  $\rho_R$ with the party $P_2$, obtains $\ket{\Phi^\pm}$ as an outcome of Bell measurement then the output state $\rho^\Phi_{1R}$ shared by parties $P_{-1}$ and $P_2$ is separable for any $\rho_1$ such that $p_1 < \frac{1}{\sqrt2}$ and for some $p$ and $\al$. Hence, for states with such parameters, if the outcome of at least one Bell measurement is $\ket{\Phi^\pm}$ there is no possibility to activate nonlocality by performing further entanglement swappings.
In Fig. \ref{ropo} we present the range of parameters $p$ and $\al$ for which $\rho^\Phi_{RR}$ and $\rho^\Phi_{1R}$ are separable.

(\textit{ii})
Without loss of generality we show that we cannot activate nonlocality between parties $P_1$ and $P_{n+1}$, where $n \geq 2$. Let us suppose that each of the parties $P_2,..., P_{n}$ obtains $\ket{\Psi^\pm}$ as a result of Bell measurement. In such a case the resulting state (after possible phase correction) is of the form
\begin{eqnarray}\label{roRn}
\rho_{R,n} = p_{R,n}  \proj{\Psi_{R,n}} + (1-p_{R,n}  )\proj{00},
\end{eqnarray}
where
\begin{eqnarray}
p_{R,n} = \frac{- p\cos2\al}{1-p-p\cos2\al+ \frac{2(p-1)}{1+ \tan^{2n}\al}} ,
\end{eqnarray}
\begin{eqnarray}
\ket{\Psi_{R,n}} =\sin \al_{n}\ket{01} + \cos \al_{n}\ket{10},
\end{eqnarray}
and
\begin{eqnarray}
\sin \al_{n} = \frac{\sin^{n}\al}{\sqrt{\sin^{2n}\al + \cos^{2n}\al}}.
\end{eqnarray}

For $\al > \pi/4$ we obtain $\al_{n}>\al$ and $p_{R,n}<p$. Hence, if the initial state $\rho_R$ does not violate CHSH inequality then the final state $\rho_{R,n}$ does not violate CHSH inequality either (see Fig \ref{roRL}).

(\textit{iii})
Let us suppose that the parties $P_{-k+1}$, $P_{-k+2}$,..., $P_{k-1}$, $P_{k+1}$,..., $P_{k+n-1}$ obtain $\ket{\Psi^\pm}$ as results of their measurement.  Hence, the parties $P_{-k}$ and $P_{k}$ will share a state (\ref{ropon}) and the parties $P_{k}$ and $P_{k+n}$ will share a state (\ref{roRn}). The latter state is of the form (\ref{roR}) and does not violate CHSH inequality. If now the party $P_{k}$ obtains $\ket{\Psi^\pm}$ as a result of his measurement then the parties $P_{-k}$ and $P_{k+n}$ will share a state
\begin{eqnarray}\label{rokn}
& \rho_{R,n,k} = \\
& p_{R,n,k}  \proj{\Psi_{R,n}} + (1-p_{n,k}  )\proj{00},\nonumber
\end{eqnarray}
with
\begin{eqnarray}
p_{R,n,k} =        \frac{p_{R,n}   (\sin^{2(n+1)}\al + \cos^{2(n+1)}\al)   }{ 1+ 2 ( \frac{1}{p_{k}} -1) p_{R,n}  \cos^{2n}\al}.
\end{eqnarray}
which is of the form (\ref{roRn}). Because $p_{R,n,k}  <  p_{R,n}$, the state $\rho_{R,n,k}$ does not violate CHSH inequality.

In (\textit{ii}) and (\textit{iii}) we did not consider the case where at least one party obtained $\ket{\Phi^\pm}$ as a result of Bell measurement because for appropriate choice of parameters the resulting state was separable.

%\section{Conclusions}\label{sec4}

In conclusion we considered activation of nonlocal correlations by performing entanglement swappings on a chain of bipartite states. In order to activate nonlocality for some states a single entanglement swapping is insufficient. In particular we have shown that before some critical number of entanglement swappings is achieved the output state does not violate CHSH inequality.

Our results generalize results derived in \cite{WMGC} and \cite{CRS} where only chains consisting of three parties were considered. In particular in \cite{WMGC} the authors considered only entanglement swapping performed by a single party. However, as we have seen even if we cannot activate nonlocality in a chain of three parties if one party performs entanglement swapping, it is possible to activate nonlocality in a chain of $n$ parties, where $n-2$ parties perform entanglement swappings. On the other hand, in \cite{CRS} more general measurements than Bell measurement by a single party were considered. Our results differ also from those derived in \cite{Ver} where only two parties were considered. It was shown there that if the parties share two entangled states and each of them does not violate CHSH inequality then the tensor product of these states can violate CHSH inequality. We also note that the effect observed in Ref. \cite{Ver}, in contrast to the effect observed in our paper, does not require postselection.

\addvspace{10pt}

\begin{acknowledgments}

We would like to thank Ryszard Horodecki for helpful discussion.
WK, MM and AG are supported by the Polish Ministry of
Science and Higher Education Grant no. IdP2011 000361.
WL is supported by the Polish Ministry of
Science and Higher Education Grant no. N202 208538 and the EU program Q-ESSENCE (Contract
No.248095).
The contribution of MM is supported within the International PhD Project
"Physics of future quantum-based information technologies"
grant MPD/2009-3/4 from Foundation for Polish Science.

\end{acknowledgments}

\end{document}